# O:H-O Bond Anomalous Relaxation Resolving Mpemba Paradox

Xi Zhang[1,2] Yongli Huang[3], Zengsheng Ma[3], Chang Q Sun[1,2,3,*]

1. NOVITAS, School of Electrical and Electronic Engineering, Nanyang Technological University, Singapore 639798
2. Center for Coordination Bond and Electronic Engineering, College of Materials Science and Engineering, China Jiliang University, Hangzhou 310018, China
3. Key Laboratory of Low-Dimensional Materials and Application Technologies (Ministry of Education) and Faculty of Materials, Optoelectronics and Physics, Xiangtan University, Hunan 411105, China
   Correspondence to: ecqsun@ntu.edu.sg (C.Q.S.) associated with honorary appointments at 2 and 3.

**Notes added:**

**Authors would like to thank the Editor and Journalists for featuring this work and numerous beneficial responses from peers, which the authors never expect.**

**This effect is a thermal "source-drain" problem subjecting to three key factors: heat emission, conduction, and dissipation.**

**Mpemba effect happens only under the critical conditions of energy rejection at a rate of history dependent, high gradient of conductivity (temperature is an indicator) in the liquid, and high gradient of temperature of the ambient. Otherwise no such effect displays as the O:H-O bond is so sensitive that any "adiabatic block" instead of "suck" of heat prevents this happening, which is why this phenomenon if hardly reproducible.**

**How about the inverse-colder water boils quicker? – Good topic to examine!**

**The following references laid foundations for this project:**

[1] Y. Huang, et al, *Size, separation, structure order, and mass density of molecules packing in water and ice*. Sci Rep, 2013. **3**: 3005.
[2] Y. Huang, et al, *Hydrogen-bond asymmetric local potentials in compressed ice*. J. Phys. Chem. B. *B* **117**, 13639-13645 (2013)
[3] C.Q. Sun, et al, *Density and phonon-stiffness anomalies of water and ice in the full temperature range*. J Phys Chem Lett, 2013. **4**: 3238-3244.
[4] C.Q. Sun, X. Zhang, and W.T. Zheng, *Hidden force opposing ice compression*. Chem Sci, 2012. **3**: 1455-1460.
[5] C. Q. Sun, *et al*. *Density, Elasticity, and Stability Anomalies of Water Molecules with Fewer than Four Neighbors*. *J Phys Chem Lett* **4**, 2565-2570 (2013).
[6] Zhang, X. & Sun, C. Q. Skin supersolidity slipperizing ice surface. *http://arxiv.org/abs/1310.0888*
[7] C.Q. Sun, *Relaxation of the Chemical Bond*, Springer Ser Chem Phys, in press.

## Abstract

We demonstrate that the Mpemba paradox arises intrinsically from the energy release whose rate depends on its amount of initial storage in the covalent H-O part of the O:H-O bond in water. Generally, heating raises the energy of a substance by lengthening and softening all bonds involved. However, heating shortens and stiffens the H-O covalent bond in water as a consequence of inter-electron-electron pair repulsion and the thermal expansion of the O:H nonbond [*J Phys Chem Lett* 4, 2565 (2013); *ibid* 4, 3238 (2013)]. Cooling shortens the O:H bond, which kicks the H-O bond from low-energy state to high-energy state to release its energy at a rate that depends on the initial storage, therefore, Mpemba effect happens. We formulated the effect in terms of the relaxation time τ. Consistency between predictions and measurements revealed that the τ drops exponentially with the initial temperature of the water being cooled, depending extrinsically on experimental conditions.

The Mpemba effect [1], named after Tanzanian student Erasto Mpemba, is the assertion that warmer water freezes faster than colder water, even though it must pass the lower temperature on the way to freezing. There have been reports of similar phenomena since ancient times, although with insufficient detail for the claims to be replicated. As indicated by Aristotle [2]: "The fact that the water has previously been warmed contributes to its freezing quickly: for so it cools sooner". Hence many people, when they want to cool water quickly, begin by putting it in the sun. Although there is anecdotal support for this paradox [3], there is no agreement on exactly what the effect is and under what circumstances it occurs.

Observations [1, 4] in Figure 1, show the following facts: a) hot water freezes faster than the cold water under the same conditions; b) the temperature θ drops exponentially with cooling duration (t) for water transiting into ice varies with experimental conditions (volume, exposure surface, etc. For example, freezing 35 °C water takes about 90 min in (a) but 35 min in (b)); c) the skin is warmer than sites near the bottom in a beaker of water being cooled. Besides, blocking heat transfer from the skin with a film of oil drastically slowed cooling. The fact that the temperature of the skin remains higher than that in the bulk of the water throughout the process of cooling is in accordance with findings that the heat capacity of the supersolid skin is higher than the bulk interior as the intramolecular H-O bond is shorter and stronger for a water molecule with fewer than four neighbors [5].

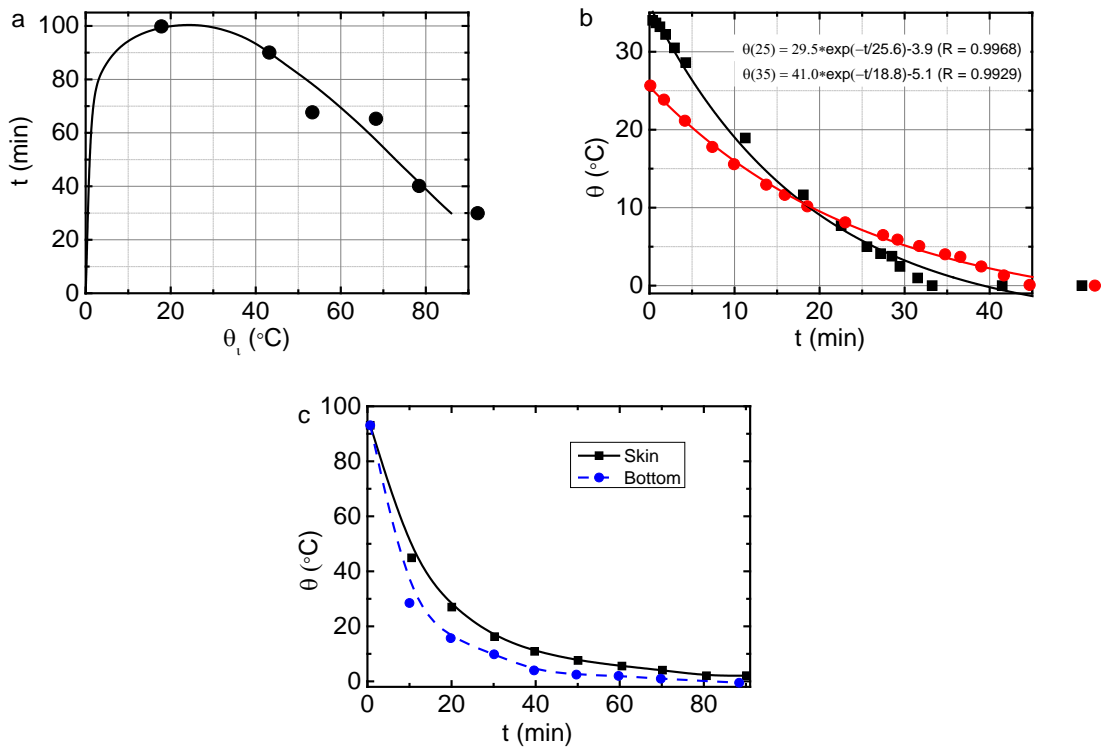

Figure 1　　　Mpemba effect. (a) Initial temperature ($\theta_i$) dependence of the cooling duration (t) for ice formation; (b) Numerical fitting of the cooling profiles (scattered data) of 30 ml water at $\theta_i$ = 35 °C and 25 °C without cover or mixing [4]; (c) time dependent skin-bottom temperature change [1].

This phenomenon forms thermodynamics paradox, but a number of possible explanations have been proposed in terms of evaporation, convection, frost, supercooling, latent heat of condensation, solutes, thermoconductivity, supercooling, etc [6-11]. Nikola Bregovićs [4], the winner of the a competition held in 2012 by the Royal Society of Chemistry calling for papers offering explanations to the Mpemba effect, explained that the effect of convection that enhances the probability of warmer water freezing first should be emphasized in order to express a more complete explanation of the effect. Even if the Mpemba effect is real, it is not clear whether the explanation would be trivial or illuminating [12]. Investigations of the phenomenon need to control a large number of initial parameters (including type and initial temperature of the water, dissolved gas and other impurities, and size, shape and material of the container, and temperature of the refrigerator) and need to settle on a particular method of establishing the time of freezing, all of which might affect the presence or absence of the Mpemba effect. The required vast multidimensional array of experiments appeared to prevent the effect from being understood. However, little attention [13] has been paid to the nature and the initial states of the water source. Why this effect happens only to water other than to other usual materials? Focusing on the relaxation dynamics of the O:H-O bond in water is necessary.

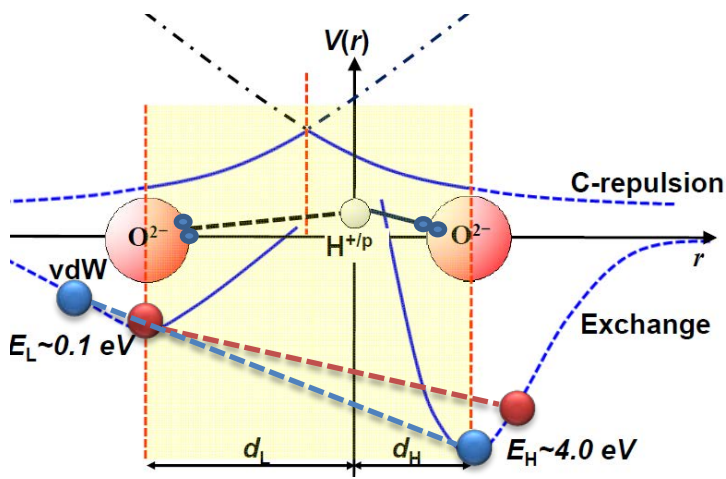

Figure 2　　　O:H-O bond in water ice. The O:H-O bond is composed of the weak O:H van der Waals

(vdW) bond in the left-hand side with vdW interaction (~0.1 eV) and the strong H-O covalent bond in the right with exchange interaction (~4.0 eV). H atom is the coordination origin. Inter-electron-pair (small paring dots on oxygen) Coulomb repulsion couples the two parts to relax in the same direction by different amounts under applied stimuli such as heating (blue spheres) or cooling (red spheres) associated with energy change along the respective potential curve. The O:H follows actively the general rule of thermal expansion in liquid water while the H-O serves as the slave to follow [14]. At cooling, the contraction of the O:H bond ejects the O atom in the H-O bond from the low-energy state to high to release energy, which differentiates water from other materials in cooling.

Let us look at the hydrogen bond (O:H-O) of water and ice first [15]. The O:H-O bond forms a pair of asymmetric, coupled, H-bridged oscillators with ultra-short-range interactions, see Figure 2 [16]. The cooperative relaxation in length and energy of the O:H-O bond and the associated energy entrapment and polarization differentiate water ice from other usual materials in the structure order and physical properties under varies stimuli [5, 14, 17]. Heating lengthens and softens of the O:H bond with energy of $10^{-2}$ eV level, which contributes insignificantly to the system energy. Thermal expansion of the O:H bond shortens and stiffens the H-O bond [14] because of the coupling by the repulsion between electron pairs on oxygen. This event results in the blue shift of the H-O stretching vibration frequency [18] and the entrapment of the O 1s binding energy [19], as given in Table 1. The heating-cooling reverses oxygen coordinates along the potential curves. The red spheres correspond to oxygen atoms in the cold state and the blue ones to the hot state. Being opposite to other usual materials, heating stores energy $\Delta E$ into the H-O covalent bond,

$$\Delta E = \eta_H \Delta \theta \propto \Delta E_H \propto \left( d_{Hi} - d_{H0} \right)^2 \equiv \Delta d_H^2$$

The $\eta_H$ is the specific heat of the H-O bond, which approximates constant in liquid phase. The energy is proportional to the vibration energy that varies with $\Delta d_H^2$. The $d_{Hi}$ and the $d_{H0}$ correspond to the H-O length at the initial and the final temperature water and the referential ice. The higher the temperature, the greater of the $\Delta d_H^2$ is.

At cooling, the shorter and stiffer H-O bond will be kicked up in the potential curve by O:H bond contraction, which releases energy to the drain at ice or supercooling state. The rate of energy release is proportional to the initial temperature of water. This process is like suddenly releasing a compressed spring at different extent of deformation with the kicking by O:H contraction as an addition of the force propelling the energy release.

In the Mpemba process, water serves as the source and the refrigerator ambient as the drain of constant temperature. The cooling environment and the cooling processes are all identical to the water at different initial temperatures without any discrimination in an experiment. Therefore, the processes and the rates of energy release from water vary intrinsically with the initial energy state of the sources.

Let us formulate the observations. Figure 1b shows that the θ decays exponentially with time required for ice formation. Therefore, one can formulate the heat releasing process in terms of:

$$\begin{cases} d\theta = -\theta \dfrac{dt}{\tau} & (decay\ function) \\ \tau^{-1} = \sum_i \tau_i^{-1} & (relaxation\ time) \end{cases}$$

and the solution,

$$\theta = \theta_i \exp\left(-\frac{\Delta t}{\tau}\right) - b$$

or,

$$\tau = -\Delta t \left[ Ln\left(\frac{\theta_f + b}{\theta_i}\right) \right]^{-1}$$

(1)

The $\tau$ is the sum of $\tau_i$ over all the possible factors of heat loss during cooling (convection, radiation, etc.), including the initial temperature of the source. It is unnecessary for one to discriminate one process from the other in the process of cooling if measurements are conducted under the same conditions. Counting of the resultant effect on the process of relaxation suffices. In estimations, an offset of the $\theta_f$ by a constant b (takes 5) ensures the theoretical curve to cross 0 °C.

Fitting to the measurements in Figure 1b yields the respective $\tau$ for the two curves, which is indeed $\theta_i$ dependent. The relaxation time $\tau$ is shorter for water cooling from higher $\theta_i$. This fact indicates that the thermal relaxation depends on the initial state of water samples under the identical ambient conditions. A comparison of Figure 1a and b shows that the relaxation time is different for water cooling from the same $\theta_i$ in different experiments. For example, cooling one drop of water and one cup of water from the same initial temperature need different times [11]. Therefore, the $\tau$ varies with experimental conditions but this extrinsic effect can be minimized by proper calibration. Nevertheless, one needs to focus on the $\theta_i$ trend of the $\tau$. Numerical calculation based on eq (1) and the fitting to the measurements in Figure 1a results in the $\theta_i$ dependence of $\tau$, as shown in Table 1 and Figure 3a. The relaxation time drops exponentially with the initial value of $\theta_i$.

The relaxation time depends intrinsically on the initial H-O bond energy. The conversion of size, separation, structure order, and mass density of molecules packing in water ice has yielded the H-O bond length at different temperatures [15]. A Lagrangian solution to the vibration of the O:H-O oscillators gives the H-O bond energy $E_H$ at different temperatures with the known H-O vibration frequency [18] and the known $d_H$ [15] as input [16]. Table 1 and Figure 3 also show the correlation between the relaxation time $\tau$ and the $\Delta d_H^2$, H-O vibration frequency $\omega_H$, and bond energy $E_H$.

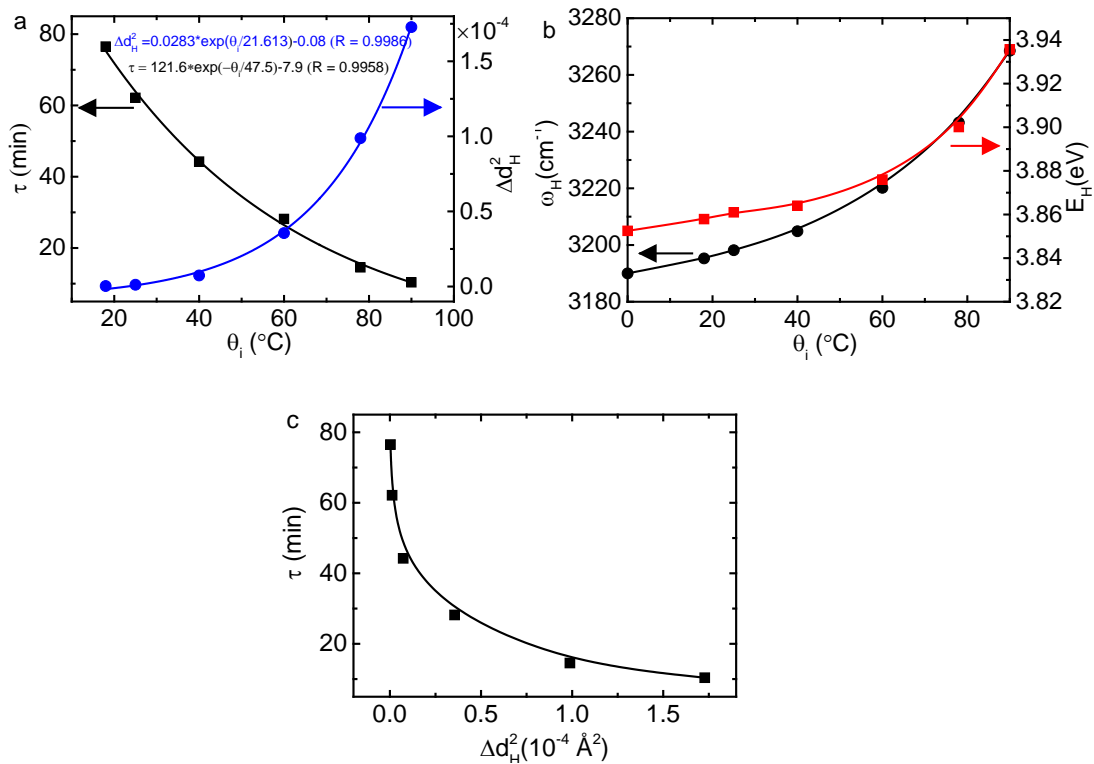

Figure 3 Temperature $\theta_i$ dependence of (a) the relaxation time $\tau$ and the square of H-O strain $\Delta d_H^2$ and (b) the vibration frequency $\omega_H$ and binding energy $E_H$ of O-H covalent bond. (c) correlation between the $\tau$ and the $\Delta d_H^2$.

Table 1 $\theta_i$ dependence of the freezing time t, relaxation time $\tau$, and the length $d_x$, vibration frequency $\omega_x$, binding energy $E_X$, in comparison with XPS O-1s core-level binding energy (x = L, H represent for the O:H and the H-O bond, respectively). $E_x$ is obtained based on the Lagrangian solution iteration with the known $d_x$ and the $\omega_x$ as input.

| $\theta_i$ (°C) | t (min) | $\tau$ (min) | H-O covalent bond | | | | | O:H vdW bond | | |
|---|---|---|---|---|---|---|---|---|---|---|
| | | | $d_H$ (Å) | $(d_H-d_0)^2$ $(10^{-4} Å^2)$ | O 1s (eV) | $\omega_H$ $(cm^{-1})$ | $E_H$ (eV) | $d_L$ (Å) | $\omega_L$ $(cm^{-1})$ | $E_H$ (eV) |
| Ref | [1] | | [15] | | [19] | [18] | | [15] | [14] | |
| 90 | 30 | 10.38 | 0.9883 | 1.728 | - | 3268.52 | 3.9357 | 1.7366 | | |
| 78 | 40 | 14.56 | 0.9915 | 0.988 | 526.90 | 3243.18 | 3.9001 | 1.7256 | | |
| 60 | 70 | 28.17 | 0.9955 | 0.355 | - | 3220.18 | 3.8760 | 1.7117 | | |
| 40 | 92 | 44.24 | 0.9987 | 0.073 | 526.80 | 3204.86 | 3.8640 | 1.7005 | | |
| 25 | 100 | 62.13 | 1.0038 | 0.012 | - | 3198.17 | 3.8610 | 1.6822 | 75 | |
| 18 | 98 | 76.51 | 1.0009 | 0.003 | 526.85 | 3195.29 | 3.8579 | 1.6926 | | |
| 5 | | | | | | | | | 175 | |
| 0 | - | - | 1.0000 | - | - | 3190.00 | 3.8525 | 1.6958 | 220 | |

We have thus formulated and verified the Mpemba paradox by introducing the relaxation time $\tau$ that depends intrinsically on the H-O bond energy at the initial temperature and extrinsically on the experimental conditions such as the volume, area of exposure surface of the samples. It is clarified that the Mpemba paradox arises intrinsically from the release of energy stored initially in the covalent H-O part of the hot O:H-O bond. Heating stores energy by shortening and stiffening the H-O covalent bond. The H-O bond releases its energy at a rate that depends exponentially on the initially stored energy and the temperature difference between the source and the drain, and therefore, Mpemba effect happens. Consistency between predictions and measurements revealed that the $\tau$ drops exponentially with the initial temperature of the water being cooled.

1.   E.B. Mpemba and D.G. Osborne, *Cool?* Phys. Educ., 1979. **14**: 410-413.
2.   Aristotle, *Meteorology* 350 B.C.E: http://classics.mit.edu/Aristotle/meteorology.1.i.html.
3.   D. Auerbach, *Supercooling and the Mpemba effect - when hot-water freezes quicker than cold.* American Journal of Physics, 1995. **63**(10): 882-885.

4. N. Bregović, *Mpemba effect from a viewpoint of an experimental physical chemist.* http://www.rsc.org/images/nikola-bregovic-entry_tcm18-225169.pdf, 2012.
5. C.Q. Sun, X. Zhang, J. Zhou, Y. Huang, Y. Zhou, and W. Zheng, *Density, Elasticity, and Stability Anomalies of Water Molecules with Fewer than Four Neighbors.* J Phys Chem Lett, 2013. **4**: 2565-2570.
6. M. Jeng, *The Mpemba effect: When can hot water freeze faster than cold?* American Journal of Physics, 2006. **74**(6): 514.
7. J.D. Brownridge, *When does hot water freeze faster then cold water? A search for the Mpemba effect.* American Journal of Physics, 2011. **79**(1): 78.
8. C.A. Knight, *The Mpemba effect: The freezing times of hot and cold water.* American Journal of Physics, 1996. **64**(5): 524-524.
9. M. Vynnycky and N. Maeno, *Axisymmetric natural convection-driven evaporation of hot water and the Mpemba effect.* Int. J. Heat Mass Transfer, 2012. **55**(23-24): 7297-7311.
10. J.I. Katz, *When hot water freezes before cold.* American Journal of Physics, 2009. **77**(1): 27-29.
11. S. Esposito, R. De Risi, and L. Somma, *Mpemba effect and phase transitions in the adiabatic cooling of water before freezing.* Physica a-Statistical Mechanics and Its Applications, 2008. **387**(4): 757-763.
12. P. Ball, *Does Hot water freeze first.* Physics world, 2006. **19**(4): 19-21.
13. L.B. Kier and C.K. Cheng, *Effect of Initial Temperature on Water Aggregation at a Cold Surface.* Chemistry & Biodiversity, 2013. **10**(1): 138-143.
14. C.Q. Sun, X. Zhang, X. Fu, W. Zheng, J.-l. Kuo, Y. Zhou, Z. Shen, and J. Zhou, *Density and phonon-stiffness anomalies of water and ice in the full temperature range.* J Phys Chem Lett, 2013. **4**: 3238-3244.
15. Y. Huang, X. Zhang, Z. Ma, Y. Zhou, J. Zhou, W. Zheng, and C.Q. Sun, *Size, separation, structure order, and mass density of molecules packing in water and ice.* Sci Rep, 2013. **3**: 3005.
16. Y. Huang, X. Zhang, Z. Ma, Y. Zhou, G. Zhou, and C.Q. Sun, *Hydrogen-bond asymmetric local potentials in compressed ice.* J. Phys. Chem. B. **DOI: 10.1021/jp407836n**.
17. C.Q. Sun, X. Zhang, and W.T. Zheng, *Hidden force opposing ice compression.* Chem Sci, 2012. **3**: 1455-1460.
18. P.C. Cross, J. Burnham, and P.A. Leighton, *The Raman spectrum and the structure of water.* J. Am. Chem. Soc., 1937. **59**: 1134-1147.
19. T. Tokushima, Y. Harada, O. Takahashi, Y. Senba, H. Ohashi, L.G.M. Pettersson, A. Nilsson, and S. Shin, *High resolution X-ray emission spectroscopy of liquid water: The observation of two structural motifs.* Chem. Phys. Lett., 2008. **460**(4-6): 387-400.